\documentclass[final, twocolumn]{elsarticle}
\usepackage{amsmath}
\usepackage{footmisc}
\usepackage{epstopdf}
\usepackage{graphicx}
\usepackage{comment}
\usepackage{xcolor,soul,framed,caption}
\colorlet{shadecolor}{yellow}

\usepackage{amssymb}

\journal{Journal of Non-Crystalline Solids}

\begin{document}

\begin{frontmatter}



\title{Universal Sound Absorption in Amorphous Solids: A Theory of Elastically Coupled Generic Blocks}


\author[label1]{Dervis C. Vural}
\author[label1,label2]{Anthony J. Leggett}
\address[label1]{University of Illinois at Urbana Champaign, 1110 W. Green Street, Urbana, IL, USA 61801}
\address[label2]{National University of Singapore, Block S 12, 2 Science Drive 3, Singapore 117542}

\begin{abstract}
 Glasses are known to exhibit quantitative universalities at low temperatures, the most striking of which is the ultrasonic attenuation coefficient $Q^{-1}$. In this work we develop a theory of coupled generic blocks with a certain randomness property to show that universality emerges essentially due to the interactions between elastic blocks, regardless of their microscopic nature.
\end{abstract}


\end{frontmatter}

\section{Introduction}
Starting with the pioneering study of Zeller and Pohl \cite{pohl71}, experimental work over the last 40 years has shown conclusively that the thermal, acoustic, and dielectric properties of virtually all amorphous materials are not only qualitatively different than those of crystals, but also show a truly amazing degree of quantitative universality\cite{anderson86,mb,pohl2002}. The theoretical interpretation of the low temperature data on amorphous materials has for the last four decades been dominated by the ``tunneling two state system'' (TTLS) model. This model gives an attractive qualitative explanation of the characteristic nonlinear effects observed in ultrasonic and dielectric absorption (saturation, echoes, hole burning). In addition it predicts a frequency and temperature dependence of the ultrasonic absorption $Q^{-1}(\omega,T)$ (here defined as $l^{-1}\lambda/2\pi^2$ in terms of the phonon mean free path and wavelength) which appears to be in fairly good agreement with the experimental data; in the present context we note that $Q^{-1}$ is predicted to be independent of $\omega$ and $T$, and the \emph{same} within a factor of 2 in two regions of the parameter space, in which, at least in terms of the model, the physics is very different, namely the high frequency ``resonance regime'' ($k_BT$, $\hbar/\tau \ll \omega$, where $\tau$ is a characteristic relaxation time of the thermally excited TTLS's) and the low frequency ``relaxation regime'' ($\omega\ll\hbar/\tau\ll k_BT$). This prediction appears to agree reasonably well with the data (see figures 2 and 3 in \cite{pohl2002}). Finally, at the cost of introducing a fairly large number of fitting parameters, the model can reproduce most of the experimental data in the low-temperature regime reasonably quantitatively.

Nevertheless, there are a number of problems with the TTLS model. First, while in a few cases (such as KBr-KCN solutions \cite{deyoreo86}) it is possible to make a plausible identification of the ``two level systems'', in most amorphous materials their nature remains a matter of conjecture. Secondly, the model as such says nothing about the behavior at intermediate temperatures (1K-30K) which also shows a very strong degree of qualitative (though not quantitative) universality. A third difficulty relates to the striking \emph{quantitative} universality and small numerical value of the quantity $Q^{-1}(\omega)$; whether observed directly or inferred (in the ``resonance regime'') from the coefficient of the $\log(T/T_0)$ term in the ultrasound velocity shift, this quantity has the value ($3\pm 2)\times 10^{-4}$ for almost all non-metallic glasses measured to date \cite{pohl2002}. While the TTLS model contains enough independent fitting parameters to ``explain'' this numerical result, the explanation requires a degree of statistical coincidence between these parameters which has no obvious basis in the model, and is prima facie nothing short of amazing. Finally, the model in its original form neglects the fact, which is emphasized below, that as a result of interaction with the strain (phonon) field, the TTLS's must acquire a mutual interaction \cite{joffrin75}; while there exist theoretical approaches which take this feature into account and even use it \cite{burin96} to attempt to account for the small universal value of $Q^{-1}$, it is not obvious that at this end of the process the TTLS structure is preserved, so that a question of self-consistency may arise. 

In \cite{leggett88}, \cite{leggett91} the conjecture was made that if we start from a very generic model in which at short length scales there is a nonzero contribution to the stress tensor from some non-phononic degrees of freedom whose only necessary feature is that their spectrum is not harmonic-oscillator-like, and take into account their phonon-mediated mutual interaction, we will recover at long length scales a picture which reproduces most, if not all, features of the experimental data below 1 K. The goal of the present paper is quite modest: To attempt a somewhat more quantitative justification of this conjecture with respect to one specific feature, namely the (near) frequency independence and small universal value of $Q^{-1}$ in the regime $k_BT,\hbar/\tau\ll\omega$ (i.e. what in the TTLS model is known as the high frequency resonance regime). We do not attempt to discuss here the behavior of $Q^{-1}$ in other regimes (except for  $\hbar/\tau\ll\omega < k_BT$), non-linear effects or (except briefly at the end of section 2.2) the intermediate-temperature behavior.

We believe that it is one of the strengths of the present work that our result does not rely on adjustable parameters, or the existence of \emph{other} microscopic (unmeasurable) universal ratios to explain the observable one \cite{yu89, wolynes01,turlakov04,parshin94,stamp09} (though cf.\cite{burin96}). The only two inputs on which our outcome depends sensitively are the ratios $c_l/c_t$ and $\chi_l/\chi_t$ (cf. below for details of the notation) both of which are observed \emph{experimentally} to vary little between different amorphous systems (also cf. Appendix-A). Our third input $r_0$, which is the size of a ``microscopic amorphous block'' (defined below) only enters into our equations logarithmically.

The organization of the paper is as follows: In section 2 we define our model and introduce the central object of our study, namely the dimensionless stress-stress correlation, whose thermally-averaged imaginary part is the measured ultrasonic absorption $Q^{-1}$. In section 3 we carry out a real-space renormalization calculation of the \emph{average} of $Q_m^{-1}(\omega)$ over the frequency $\omega$ and the starting state $m$ (for details of the notation see below) and show that it vanishes logarithmically with the volume of the system and, for experimentally realistic volumes, has a surprisingly small value, $\sim0.015$. In section 4 we argue on the basis of a calculation up to second order in the phonon induced interaction, that the functional form of $Q^{-1}(\omega)$ at $T=0$ should be ($\ln\omega)^{-1}$, and that when we combine this result with that of section 3, the numerical value of $Q^{-1}$ for experimentally relevant frequencies should be universal up to logarithmic accuracy and numerically close to the observed value $3\times10^{-4}$. In section 5 we assess the extent to which our calculations are consistent with experiments in the (linear) resonance regime. In section 6 we attempt to assess the significance of our results.

Throughout this paper we set $\hbar=k_B=1$. The notation $a$ denotes a ``typical'' atomic length scale. The suffix $\alpha=l,t$ denotes the phonon polarization ($l$=longitudinal, $t$=transverse). 

\section{Formulation of the Problem}

Consider a cube of an arbitrary isotropic amorphous material, with side L which is assumed large compared to ``microscopic'' lengths $a$ such as the typical interatomic distance, but is otherwise arbitrary. We define for such a block the strain tensor $e_{ij}$ in the standard way: If $\vec{u}(\vec{r})$ denotes the displacement relative to some arbitrary reference frame of the matter at point $\vec{r}$, then
\begin{align}\label{2.1.1}
e_{ij}=\frac{1}{2}\left(\frac{\partial u_i}{\partial x_j}+\frac{\partial u_j}{\partial x_i}\right)
\end{align}
(Note that the anti-symmetric part of the tensor $\partial u_i/\partial x_j$ corresponds to a local rotation; since a spatially uniform rotation costs no energy, any terms in the Hamiltonian associated with this part will be of order higher than zeroth in the spatial gradients, and hence for the purposes of the ensuing argument irrelevant in the renormalization-group sense; we therefore  neglect any such terms in the following).

Let us expand the Hamiltonian of the block in a Taylor series in the strain $e_{ij}$:
\begin{align}\label{2.1.2}
\hat{H}=\hat{H}_0+\sum_{ij}e_{ij}\hat{T}_{ij}+\mathcal{O}(e^2)
\end{align}
where the stress tensor $\hat{T}_{ij}$ is defined by
\begin{align}\label{2.1.3}
\hat{T}_{ij}=\partial \hat{H}/\partial e_{ij}
\end{align}
Note that in general, in a representation in which $\hat{H}_0$ is diagonal, $\hat{T}_{ij}$ will have both diagonal and off-diagonal elements. 

As usual, we can define the static elasticity modulus $\chi^{(0)}$, a fourth order tensor, by 
\begin{align}\label{2.1.4}
\chi^{(0)}_{ij:kl}\equiv V^{-1}(\partial\langle\hat{T}_{ij}\rangle/\partial e_{ij})_{eq}\nonumber\\
\equiv V^{-1}\langle\partial^2\hat{H}/\partial e_{ij}\partial e_{kl}\rangle_{eq}
\end{align}
where $V=L^3$ is the volume of the block and the suffix ``eq'' denotes that the derivative is taken in the thermal equilibrium state (both sides of eqn(\ref{2.1.3}) are implicitly functions of temperature $T$). Since by definition the properties of an isotropic amorphous material must be invariant under overall rotation, symmetry considerations constrain $\chi_{ij:kl}^{(0)}$ to have the generic form 
\begin{align}\label{2.1.5}
\chi^{(0)}_{ij:kl}=(\chi_l-2\chi_t)\delta_{ij}\delta_{kl}+\chi_t(\delta_{ik}\delta_{jl}+\delta_{il}\delta_{jk})
\end{align} 
where $\chi_l$ and $\chi_t$ are the standard longitudinal and shear elastic constants; in the approximation of an elastic continuum, these are related to the velocities $c_l$ and $c_t$ of the corresponding longitudinal and transverse sound waves (of wavelength $\lambda$ such that $a\ll\lambda\ll L$) by\footnote{Note that despite the notation $\chi$ has the characteristics of a ``stiffness'' ($\sim$ inverse susceptibility) rather than a ``susceptibility''}
\begin{align}\label{2.1.6}
\chi_{l,t}=\rho c_{l,t}^2
\end{align}
where $\rho$ is the mass density of the material. Such an approximation however throws away all the effects of interest to us, as we shall now see. 

\subsection{The Stress-Stress Correlation Function}
Let us separate out from the Hamiltonian, the purely elastic contribution $\hat{H}_{el}$, namely,
\begin{align}\label{2.2.1}
 \hat{H}_{el}(e_{ij})=\mbox{const.}+\int \frac{1}{2}d^3r&\sum_{ijkl}\chi_{ij:kl}^{(0)}e_{ij}(\vec{r})e_{kl}(\vec{r})\nonumber\\+\frac{1}{2}&\sum_i\rho\dot{\vec{u}}_i^2(\vec{r})
\end{align}
(where it is understood that the velocity is slowly varying over distances $a$, as above). Similarly we define the ``elastic'' contribution to the stress tensor $\hat{T}_{ij}$ by
\begin{align}\label{2.2.2}
 \hat{T}_{ij}^{(el)}\equiv\sum_{ijkl}\chi_{ij:kl}^{(0)}e_{kl}
\end{align}
(In above, $e_{ij}$ (and $u_i$) should strictly be treated as operators, but we prefer not to complicate the notation unnecessarily). Then quite generally, we have
\begin{align}\label{2.2.3}
 \hat{H}(e_{ij})\equiv \hat{H}_{el}(e_{ij})+\hat{H}'(e_{ij})
\end{align}
where for the moment the ``non-phonon'' term $H'(e_{ij})$ is completely general (in particular we do not assume it is necessarily ``small'' compared to $H_{el}$). In analogy to (\ref{2.1.2}) and (\ref{2.1.3}) we can define a ``non-phonon'' contribution to the stress tensor by 
\begin{align}
 \hat{H}'&=\hat{H}'_0+\sum_{ij}e_{ij}\hat{T}_{ij}'+\mathcal{O}(e^2)\label{2.2.4}\\
\hat{T}_{ij}'&=\partial \hat{H}'/\partial e_{ij}\label{2.2.5}
\end{align}
From now on we shall always take the quantities $\hat{H}_0$ and $\hat{T}_{ij}$ to refer to the non-phonon contributions and accordingly omit the primes. Note carefully that the origin of $e_{ij}$ in (\ref{2.2.4}) is not specified, and in particular it may have contributions from the phonon field.

We can now define the quantity which will be the central object of our study in this paper, namely the (non-phonon) stress-stress correlation function (linear response function) at different scales $L$. Consider an externally imposed infinitesimal sinusoidal strain field, 
\begin{align}\label{2.2.6}
e_{ij}(\vec{r},t)=e_{ij}\mbox{exp}\{i(\vec{q}.\vec{r}-\omega t)\}+\mbox{c.c} 
\end{align}
($e_{ij}$ real). This will give rise to a corresponding response of $\langle T_{ij}\rangle$:
\begin{align}\label{2.2.7}
\langle T_{ij}\rangle(\vec{r},t)=\langle T_{ij}\rangle\mbox{exp}\{i(\vec{q}.\vec{r}-\omega t)\}+\mbox{c.c}
\end{align}
where $\langle T_{ij}\rangle$ is in general complex. Then we can define the complex response function $\chi_{ij:kl}(\vec{q},\omega)$ in the standard way\footnote{For brevity we omit some technical complications involved in the precise definition of $\chi$; these are very standard, see e.g. \cite{pines}}
\begin{align}\label{2.2.8}
 \chi_{ij,kl}(\vec{q},\omega)=\frac{\partial \langle T_{ij}\rangle}{\partial e_{kl}}(\vec{q},\omega)
\end{align}

We will usually omit the explicit $\vec{q}$-dependence of $\chi$; it should be remembered that in general $\vec{q}$ and $\omega/c_{l,t}$ are independent variables, cf.footnote \footref{fn:note7} below.

In practice we shall usually be interested in values of $|q|$ which are close to $\omega/c_{l,t}$ and will therefore usually omit the explicit $\vec{q}$ dependence of $\chi$ (However cf. footnote \footref{fn:note7} below). 

Since for the purposes of the argument below, we shall be interested, at a geometrical scale $L$, in values of $|q|$ which are of the order of $L^{-1}$, it is not immediately obvious that the symmetries of $\chi_{ij:kl}(\vec{q},\omega)$ allow its representation in the simple form analogous to (\ref{2.1.5}); however, since it is clear that any complications associated with this consideration are sensitive to our arbitrary choice of block shape, we will assume that a more rigorous (q-space) calculation will get rid of them, and thus assume that $\chi_{ijkl}(\omega)$ can indeed be represented in the form (\ref{2.1.5}), thereby defining ``longitudinal'' and ``transverse'' response functions $\chi_{l,t}(\omega)$.

All the above considerations are independent of the scale $L$ of the block considered, provided only that this is large compared to atomic scales $a$. Let us now for a moment specialize to values of $L$ of the order of the wavelength of the phonons studied directly in glasses. In view of the small values of $Q^{-1}$ and related quantities observed experimentally in this regime, it is very plausible to assume that the coupling between the phonon and non-phonon degrees of freedom (part of the second term in (\ref{2.2.4})) is a ``weak'' perturbation on the phonon dynamics as described by  $\hat{H}_{el}$. With this assumption it is straightforward to calculate the dimensionless ultrasonic attenuation $Q_\alpha^{-1}(\omega)$ of a phonon of frequency $\omega$. Omitting the details of the derivation, we just quote the result
\begin{align}\label{2.2.9}
 Q_\alpha^{-1}(\omega)\equiv\lambda l^{-1}/2\pi^2=\mbox{Im}\chi_\alpha(\omega)/(\pi\rho c_\alpha^2)
\end{align}
where the $4^{th}$ rank tensor quantity $\mbox{Im}\chi(\omega)$ (which we shall need in full generality below) is given explicitly, in the representation in which $\hat{H}_0$ is diagonal, by the formula
\begin{align}
\mbox{Im}\chi_{ij:kl}(\omega)=&\sum_mp_m\chi^{(m)}_{ij:kl}(\omega)\label{2.2.10}\\
\chi_{ij:kl}^{(m)}(\omega)=L^{-3}\pi&\sum_n\langle m|T_{ij}|n\rangle\langle n|T_{kl}|m\rangle\nonumber\\
&\times\delta(E_{n}-E_{m}-\omega)\label{2.2.11}
\end{align}
where $m, n$ denote exact many-body eigenstates of $\hat{H}_{0}$, with energies $E_m$, $E_n$, and $p_m$ is the probability of occurrence of initial state $m$ (in thermal equilibrium we of course have $p_m=Z^{-1}\exp\{-\beta E_m\}$, where $Z$ is the partition function).

Since the above formulation is very generic, it is clear that the standard TTLS  model must be a special case of it, specified by particular choices of the matrix elements of $\hat{H}_0$ and $\hat{T}$. Without investigating these choices in detail (cf. below), let us note that according to this model the form of $Q_\alpha^{-1}(\omega)$ is 
\begin{align}\label{2.2.12}
 Q_\alpha^{-1}(\omega)=Q_{hf}^{-1}\mbox{tanh}(\omega/2T).
\end{align}

This form seems to agree reasonably well with experiment, with $Q_{hf}^{-1}\sim3\times10^{-4}$. Note however that if (\ref{2.2.12}) is valid for all $T$, then in view of the Kramers-Kronig formula relating $Q_{hf}^{-1}$ to $\chi_{0\alpha}$, the latter quantity actually diverges logarithmically in the limit $T\to0$. This might suggest (though it does not of course prove) that the true $T=0$ form of $Q_\alpha^{-1}$ is actually a weakly decreasing function of decreasing $\omega$; we will provide some evidence for this conjecture in section 4.

\subsection{Phonon-Induced Stress-Stress Interaction and Real-Space Renormalization Program}

Consider a whole set of blocks, each of size $L$ and described by the generic Hamiltonian (\ref{2.2.4}); for the moment we work at zero temperature. Since the quantity $e_{ij}$ contains a contribution from the phonon field, it is clear that exchange of virtual phonons will give rise to an effective (``RKKY''-type) coupling between the stress tensors of two different blocks 1 and 2, and since the phonon system by itself is harmonic, this will have the generic form
\begin{align}\label{2.3.1}
 H_{int}^{(12)}\!=\!\frac{1}{2}\!\int^{v_1}\!\!\!\!\!d\vec{r}\int^{v_2}\!\!\!\!d\vec{r}'
\sum_{ijkl}\Lambda_{ijkl}(\vec{r}-\vec{r}')T_{ij}(\vec{r})T_{kl}(\vec{r'}).
\end{align}
The function $\Lambda_{ijkl}(\vec{r}-\vec{r}')$ is calculated in the paper of Joffrin and Levelut\cite{joffrin75}; while the paper is formulated explicitly in the language of the TTLS model, it is clear that their form is generic. In the limit of large $|\vec{r}-\vec{r}'|$ (see below for the meaning of ``large'') it has the following form \cite{esquinazi-rev}, where $\vec{n}$ is the unit vector along $\vec{r}-\vec{r}'$,
\begin{align}
\Lambda_{ijkl}(\vec{r}-\vec{r}')=\frac{1}{\rho c^2_t}\frac{1}{2\pi|\vec{r}-\vec{r}'|^3}\tilde{\Lambda}_{ijkl}(\vec{n})\label{2.3.2}\\
\tilde{\Lambda}_{ijkl}=-(\delta_{jl}-3n_jn_l)\delta_{ik}+(1-c_t^2/c_l^2)[-\delta_{ij}\delta_{kl}\nonumber\\-\delta_{ik}\delta_{jl}-\delta_{il}\delta_{jk}
+3(\delta_{ij}n_kn_l+\delta_{ik}n_jn_l+\delta_{il}n_kn_j\nonumber\\+\delta_{jk}n_in_l+\delta_{jl}n_in_k+\delta_{kl}n_in_j)-15n_in_kn_kn_l]\label{2.3.3}
\end{align}
An approximation which should be adequate for our purposes is to replace $\vec{r}-\vec{r}'$ in (\ref{2.3.1}) by $\vec{R}_1-\vec{R}_2$ when $\vec{R}_s$ denotes the center of block $s$, and that $\int_{v_s}\hat{T}_{ij}(\vec{r})d\vec{r}$ is the \emph{uniform} stress tensor of the block. Then the total Hamiltonian of the $N$ coupled blocks is, in an obvious notation,
\begin{align}\label{2.3.4}
 \hat{H}_N=\sum_{s=1}^N\hat{H}_0^{(s)}\!\!+\frac{1}{2}\!\!\sum_{\substack{s,s'=1\\s\neq s'}}^N\sum_{ijkl}\Lambda_{ijkl}(\vec{R}_s-\vec{R}_s')T_{ij}^{(s)}T_{kl}^{(s')}
\end{align}
Eqn(\ref{2.3.4}) then represents the Hamiltonian $H_0$ of the ``super block'' (of side $\sim N^{1/3}L$) composed by the $N$ blocks of side $L$; we can then define the stress tensor $\hat{T}_{ij}$ for this super block and iterate the procedure until we reach the experimental scale. Note that because of the factor proportional to $L^{-3}$ in $\Lambda_{ijkl}$ and the fact that the correlation functions of the $T_{ij}$ are defined intensively to have the same factor of $L^{-3}$, the procedure is scale invariant (in 3 space dimensions) and we might hope that it will iterate to a fixed point at large length scales. 

The program we would ideally like to implement, therefore, is to start with given forms of $\hat{H}_0$ and $\hat{T}_{ij}$ at some ``short'' length scale, introduce interaction between the corresponding ``small'' blocks according to (\ref{2.3.4}), diagonalize the Hamiltonian $\hat{H}_N$, obtain the corresponding stress tensor $\partial \hat{H}_N/\partial e_{ij}^{(N)}$ and iterate the procedure up to the experimental length scales. In particular, we would be interested in the extent to which the ``output'' forms of $H_N$ and $T_{ij}$ at the experimental scale, are independent (or not!) of the ``input'' forms at the starting length scale. In practice, a meaningful implementation of this problem appears to require massive computational resources, so in the rest of this paper we will concentrate on a few results which can be plausibly obtained by analytical techniques. 

First, however, we need to discuss the question of an appropriate choice of $L$ for the ``input'' blocks. It is well known that the temperature dependence of the thermal conductivity of amorphous materials changes at a temperature of the order of $1K$; in fact, most such materials show a pronounced ``plateau'' extending very roughly between 1 and 10K. At 10K the dominant phonons have wavelengths of the order of $50$\AA, and in \cite{leggett91} it is argued that this is just the scale at which we get a crossover from ``Ising'' to ``Heisenberg'' behavior (formally, at smaller scales the approximations used in obtaining the simple $R^{-3}$ form of $\Lambda$, eqn(\ref{2.3.2}) breaks down). Thus, we tentatively take the ``starting'' block size, which we will denote $r_0$, to be $\sim50$\AA (which is still comfortably greater than $a$). Since the results obtained in sections 3 and 4 depend only logarithmically on $r_0$, they will not be particularly sensitive to this choice.

\section{Average value of $Q^{-1}$}

To introduce this section let us consider the following heuristic argument (cf. \cite{fisch80,leggett91}): Imagine a set of spin-like objects (in three spatial dimensions) which do not have direct mutual interactions but do interact with the phonon field via a term proportional to the strain, with some coefficient $\gamma$. As a result of this interaction, we will get an effective ``spin-spin'' interaction which is schematically of the form $g/r^3$ times some angular factor which is of no great interest in this context, where $g\equiv\eta\gamma^2/(\rho c^2)$ with $\eta$ a dimensionless number of order 1 (cf. below).

Assuming that for the resulting effective Hamiltonian it is possible to define some kind of single-excitation density of states $\bar{P}$ which is to a first approximation independent of $E$, it then follows on dimensional grounds that $\bar{P}$ must be of the form $\mbox{const.} g^{-1}$. Now in this model, the dimensionless  ultrasonic absorption $Q^{-1}$ is simply $(\pi/2)(\gamma^2/\rho c^2)\bar{P}$, so it follows that $Q^{-1}$ is ``universal'' (i.e. independent of $\gamma$, $\rho$ and $c$). Moreover, since all phonon ``colors'' (modes) contribute to $g$, while only one color is absorbed at a time, it follows that $Q^{-1}\sim\eta^{-1}\sim n^{-1}$ when n is the number of phonon ``colors''.

What we would like to do in this section is to try to (a) generalize this argument to a more generic model which does not necessarily assume ``single-particle'' excitations (with or without a constant density of states) (b) take into account quantitatively the existence not only of different phonon ``colors'' but of different stress tensor components, and (c) argue that steps (a) and (b) alone lead to a surprisingly small value of $Q^{-1}$ (though not small enough to explain by themselves the experimental data)

To this end, it is convenient to define for a block of size $L$ quantities $\chi_\alpha^{(m)}(\omega)$ in terms of $\chi_{ijkl}^{(m)}(\omega)$ ($\alpha=l,t$) in a way exactly analogous to that done in (\ref{2.1.5})\footnote{It may be objected that the individual $\chi_{ijkl}^{(m)}(\omega)$ do not necessarily have this symmetry; however, since all physically relevant quantities involve averages over $m$, we shall neglect this complication.}, and further define an average $\bar{\chi}_\alpha$ of $\chi_\alpha^m(\omega)$ over both frequency $\omega$ and initial state $m$, and an associated dimensionless quantity $\bar{Q}^{-1}_\alpha$ as follows:
\begin{align}
\bar{\chi}_\alpha&=N_s^{-1}U_0^{-1}\sum_m\int_0^{U_0}d\omega\chi_\alpha^{(m)}(\omega-E_m)\label{3.2.1}\\
\bar{Q}_\alpha^{-1}&=(\pi\rho c^2_\alpha)^{-1}\bar{\chi}_\alpha\label{3.2.2},
\end{align}
where $N_s$ is the number of levels below the cutoff $U_0$. The choice of $U_0$ and the range of the sum over $m$ will be specified below.

We now consider two statistically identical blocks 1 and 2, and write the complete Hamiltonian $\hat{H}$ of the coupled system in the form
\begin{align}\label{3.2.3}
 \hat{H}=\hat{H_0}+\hat{V}
\end{align}
with $H_0$ and $\hat{V}$ defined by the relevant special case of eqn(\ref{2.3.4}). Since by hypothesis the matrix elements of $\hat{H}_0^{(s)}$ and $\hat{T}_{ij}^{(s)}$ (and thus of $\hat{H}_0$ and $\hat{V}$) are uncorrelated, we may write
\begin{align}\label{3.2.4}
 \mbox{Tr}\hat{H}^2-\mbox{Tr}\hat{H_0}^2=\mbox{Tr}\hat{V}^2
\end{align}
provided, of course, that the three traces are over the same manifold of the complete many-body Hilbert space. The question of the optimal choice of this manifold is a delicate one. We know that for energy scales $E_{n,m}> U_0=hc/L$ and length scales $|R|< L$, the assumptions that give rise to the simple form in eqn(\ref{2.3.2}) no longer hold, and $\langle m|\hat{V}(\vec{R})|n\rangle$ becomes a rapidly oscillating function of $m, n, |R|$. We handle these oscillatory effects by assuming that the interactions have no effect on $E_n>U_0$, and the corresponding $|n\rangle$. Thus we take the relevant manifold to be simply spanned by those (many-body) eigenstates of $\hat{H}_0^{(s)}$ whose energies $E_0$ are less than $U_0$. By restricting ourselves within a low energy manifold, we are implicitly assuming that the interaction matrix elements that mix in high-energy states into the manifold are negligible. To be quite explicit,we \emph{assume} that the matrix element $\hat{V}_{mn}$ is negligible whenever one of the eigenenergies $m,n$ of $\hat{H}_0$ is less than $U_0$ and the other greater; call this assumption (which we will carry over to subsequent iterations of the renormalization process) assumption (A). It seems unlikely that this approximation will influence the \emph{low-energy} predictions that follow. In addition, we will see that regardless of the length scale, $U_0$ drops out of the calculation (cf. eqn(\ref{3.2.11}), and we need not worry about its precise value or $L$ dependence. We hope to relax assumption (A), and possibly others in a future study.


On general grounds, we may assume that the density of eigenvalues $E_n$ of the interacting and noninteracting systems have the generic many-body form
\begin{align}\label{2.4.1}
 p(E)=\sum_{n=0}^{\infty}a_nE^{n}
\end{align}
with dominating powers much larger than unity (in general, the dominating exponent of an extensive many-body system must be proportional to the number of particles). While the actual many-body density of states we deduce from the specific heat data, 
\begin{align}\label{2.4.11}
p(E)= p_0\exp\{\mbox{const.}(NE)^{1/2}\}                                                                                                                              
\end{align}
is consistent with this requirement, as we will see, none of our results will be sensitive to the precise choice of $a_n$. Because of the high power in eqn(\ref{2.4.1}), and taking into account that the number of states in the superblock $1+2$ is $N_s^2$, we can evaluate $\mbox{Tr}H_0^2\approx N_s^2U_0^2$ (cf. Appendix-B). Moreover, if $H^2$ is diagonalized within the manifold, it will have some maximum eigenvalue $U$, and by the same reasoning we expect $\mbox{Tr}\hat{H}^2\approx N_s^2U^2$. Thus,
\begin{align}\label{3.2.5}
 \mbox{Tr}\hat{H}^2-\mbox{Tr}\hat{H_0}^2=N_s^2(U^2-U_0^2)
\end{align}
What of the RHS of (\ref{3.2.4})? Here, there is a tricky point. We can evaluate the trace either in the eigenbasis of $\hat{H}_0$ or that of $\hat{H}$. In the first case, it is clear that  $Tr\hat{V}^2$ is proportional to $N_s^2U_0^2\left(\bar{Q}_0^{-1}\right)^2$ (where we now add the suffix 0 to indicate that the quantity in question refers to the original size-$L$ blocks); omitting the details of the algebra (cf. Appendix-B), we simply quote the result
\begin{align}\label{3.2.6}
 \mbox{Tr}V^2=CKN_s^2U_0^2\left(\bar{Q}_0^{-1}\right)^2
\end{align}
where we have defined a geometrical factor $C\equiv V^2/[16\pi^2(R_1-R_2)^6]$, which for two side-by-side cubes is $9/16\pi^2$, and where the factor $K$ is given by
\begin{align}\label{3.2.7}
 K=(8/3)[-3+4p+16q(q+p+qp-1)]
\end{align}
which for typical (experimental) values of $q\equiv1-c_t^2/c_l^2\approx0.6$ (cf. Appendix-1 for a theoretical justification) and $p\equiv\mbox{Im}\chi_l/\mbox{Im}\chi_t\approx2.6$, is $122$. We note that the $\bar{Q}_0^{-1}$ in (\ref{3.2.6}) actually refers to the transverse coefficient $\bar{Q}_{0t}^{-1}$, the suffix of which we drop for notational simplicity.

It is also possible to evaluate $\mbox{Tr}\hat{V}^2$ in the eigenbasis of $\hat{H}$, in terms of the ``renormalized'' $\chi$'s of the individual blocks 1 and 2 defined by taking, in the definition (\ref{2.2.11}), the energies $E_m$ and states $|m\rangle$ to be the eigenvalues and eigenstates of $\hat{H}$ rather than $\hat{H}_0$ (recall that thanks to assumption (A), $\hat{H}$ can be diagonalized within our submanifold). It is clear that all the geometrical factors, etc., are identical in the two cases, so we get
\begin{align}\label{3.2.8}
 \mbox{Tr}\hat{V}^2=CKN_s^2U^2\left(\bar{Q}_r^{-1}\right)^2
\end{align}
where $\left(\bar{Q}_r^{-1}\right)$ is the averaged absorption of one of the two blocks. We now face the difficulty that this is \emph{not} necessarily the averaged absorption of the superblock 1+2 (or even related to it by a simple numerical factor), because the definition of the latter involves the squared matrix elements of the \emph{total} stress tensor of the superblock, $(\hat{T}_{ij}^{(1)}+\hat{T}_{ij}^{(2)})$ and thus contains terms like $\langle n|\hat{T}_{ij}^{(1)}|m\rangle\langle m|\hat{T}_{ij}^{(2)}|n\rangle$ (where $|m\rangle, |n\rangle$ now denote eigenstates of $\hat{H}$); while such terms were originally (in the absence of $\hat{V}$) uncorrelated, it is not obvious that they remain uncorrelated after $\hat{V}$ is taken into account. We shall, however, argue that on average those terms are likely to be small compared to terms of the form $|\langle m|T_{ij}^{(s)}|n\rangle|^2$, because $\hat{V}$ involves \emph{all} tensor components of $\hat{T}^{(1)}$ and $\hat{T}^{(2)}$ while the correlation only involves the \emph{same} component of $\hat{T}^{(1)}$ and $\hat{T}^{(2)}$;call this assumption (B). If this argument is accepted, we can identify the $Q_r^{-1}$ in (\ref{3.2.8}) with the physical inverse absorption of the superblock, which we denote simply $Q^{-1}$.

Thus, putting together eqns.(\ref{3.2.5}),(\ref{3.2.6}) and (\ref{3.2.8}), we have
\begin{align}\label{3.2.9}
 N_s^2(U^2-U_0^2)&=CKN_s^2U_0^2\left(\bar{Q}_0^{-1}\right)^2\nonumber\\&=CKN_s^2U^2\left(\bar{Q}^{-1}\right)^2
\end{align}
from which we can express $\bar{Q}^{-1}$ (the inverse absorption of the superblock 1+2) in terms of that, $\bar{Q}_0^{-1}$ of the original blocks 1 or 2:
\begin{align}\label{3.2.10}
 \bar{Q}^{-1}=[1/(\bar{Q}_0^{-1})^2+CK]^{1/2}
\end{align}
We can now repeat this procedure, by combining the superblock 1+2 (which we recall is the two cubes 1 and 2 side by side) with a similar superblock 3+4, and finally combining the square structure so formed with a similar one to form a cube of side $2L$. The only difference with the operation carried out explicitly above lies in the geometrical factor $C$, which is 4 times larger than that in (\ref{3.2.6}) for step 2 and 16 times larger for step 3\footnote{To be sure, the approximation made in going from (\ref{2.3.1}) to (\ref{2.3.4}) is somewhat less reliable than when combining two cubes side by side, and different ways of treating the sum (integral) can alter $C$; here, as at several other points, it would seem that a more reliable estimate of the numerical factors would require an explicit k-space calculation.}. As a result, the $\bar{Q}^{-1}$ of the cube of size $2L$ is related to that of the original size-$L$ one by (now denoted $Q_L$ for clarity)
\begin{align}\label{3.2.11}
 \bar{Q}_{(2L)}^{-1}\!\!=\!\!\left[\frac{1}{(\bar{Q}_{(L)}^{-1})^2}+\frac{189}{16\pi^2}K\right]^{-1/2}\!\!\!\!\!\!\!=\!\left[\frac{1}{(\bar{Q}_{(L)}^{-1})^2}+K_0\right]^{-1/2}
\end{align}
with $K_0\sim150$. This completes the first stage of the iteration: We see that even in a single step of the iteration the effects of the phonon-induced stress-stress interaction is to strongly renormalize the average ultrasonic absorption downwards.

We now consider the effect of iterating the step which led to (\ref{3.2.11}), by combining eight cubes of size $2L$ to make one of size $4L$; for convenience we keep the original definition of the low-energy manifold ($E_m<\omega_0$), though other choices are also possible. Since the only point at which $\omega_0$ actually enters the result is implicitly in the definition of the ``average'' in $\bar{Q}^{-1}$ the scale invariant nature of the problem implies that all considerations are exactly the same as at the first stage, and we simply recover (\ref{3.2.11}) with the replacement of $L$ by $2L$ and $2L$ by $4L$. Continuing the iteration up to a spatial scale $R$, we find
\begin{align}\label{3.2.12}
\bar{Q}^{-1}(R)=\left[\bar{Q}_{(L)}^{2}+K_0\log_2(R/L)\right]^{-1/2}
\end{align}
with the constant $K_0$ weakly material dependent and approximately $150$. It is remarkable that this formula predicts that at $T=0$ the ultrasonic absorption completely vanishes in the thermodynamic limit. While we know of no particular reason why this behavior must be unphysical, in practice we would guess that for any finite ultrasound wavelength $\lambda$, $R$ would be replaced by a quantity of order $\lambda$\footnote{To avoid confusion, we shall emphasize that in general (e.g. in considering the average over $\omega$ in (\ref{3.2.12})) the inverse length scale $q\equiv R^{-1}$ should be regarded as a different variable than $\omega/c$; it is only when evaluating the predictions for the experimentally measured value of $Q^{-1}$ that we should take $q=\omega/c$\label{fn:note7}}. For experimentally realistic values $R$ and the choice of $L\sim50$\AA (cf. end of section 2) the value of $\bar{Q}^{-1}$ is approximately $0.015$\footnote{Or slightly larger if we take $R\sim\lambda$}. While $\bar{Q}^{-1}$ is material independent, its value is rather large compared to the experimental value in the MHz-GHz range. We think this is due to the contribution to the average, of the rapid increase in $Q^{-1}$ at higher frequencies, as manifest in the thermal conductivity data around 10K. To obtain the experimentally observed absorption in the MHz-GHz range we need further considerations, one of which we now explore in section 4. 

\section{The Frequency-Dependence of $\chi(\omega)$: Second Order Perturbation Calculation}
In the last section we obtained an expression (\ref{3.2.12}) for the value of $Q^{-1}(\omega)$ (or equivalently $\chi(\omega)$) averaged over the specified (low-energy) initial states $m$ and over a frequency range $U$, but we could deduce nothing about the frequency-dependence of $\chi$. In the present section we shall attempt to deduce some conclusions about the frequency-dependence on the basis of the specific ansatz (\ref{2.4.2}) concerning the form of the ``input'' $\chi$ at scale $L$. The reader should be warned that the argument of this section is heuristic and somewhat unorthodox.

We consider the consequences (at $T=0$) for $\chi(\omega)$ on taking into account the interaction term in $\hat{V}$ in (\ref{2.3.4}) up to second order in perturbation theory for two neighboring blocks 1 and 2. In general there will be changes in $\chi(\omega)$ (eqn (\ref{2.2.11})) due to (a) modification of the many-body density of states (DOS) and (b) modification of the matrix elements of the stress tensors $T_{ij}$ (which, obviously, we need to calculate for the superblock 1+2, not for blocks 1 and 2 individually). An important assumption which we shall make in this section is that, at least for the purposes of obtaining qualitatively correct results, \emph{effect (b) may be neglected,}\footnote{One of us has in fact calculated \cite{dcv-future} the modification of $\chi(\omega)$ to second order in $V$; the expression is an extremely messy sum of a large number of terms with different signs, and (unsurprisingly) its quantitative evaluation requires assumptions on the matrix elements of $T_{ij}$ which go beyond (\ref{2.4.2})} so that all we need to do is to calculate the shift due to $\hat{V}$ in the energies $E_n$ of the superblock. The first order term vanishes because of the lack of correlation in the matrix elements $T_{ij}^{(1)}$ and $T_{ij}^{(2)}$; the second order term can be expressed using eqns (\ref{2.2.11}) and (\ref{2.4.2}) (when $\Delta$ is the correction to the energy of a superblock state $E_n=E_{n_1}+E_{n_2}\equiv \omega$)
\begin{align}\label{4.1.1}
\Delta(\omega)=CK\!\int_{-E_{n_1}}^\infty\!\!\int_{-E_{n_2}}^\infty\!\!\!Q^{-1}_{0,n_1}(\omega')Q^{-1}_{0,n_2}(\omega'')\nonumber\\\times\theta(U-|\omega'+\omega''|)(-\omega'-\omega'')^{-1}d\omega'd\omega''
\end{align}
where $C$ and $K$ are the quantities defined above in connection with eqn(\ref{3.2.7}). To evaluate this formula we need an input form for $Q_0^{-1}$. For this we shall make the simplest possible choice which is consistent with our general assumptions, namely the ``random'' form.
\begin{align}\label{2.4.2}
 \chi_{\alpha}^{(m)}(\omega)/(\pi\rho c_\alpha^2)&=\mbox{const.}\theta(E_m+\omega)\nonumber\\
&\equiv Q_0^{-1}\theta(E_m+\omega)
\end{align}
Note that eqn(\ref{2.4.2}) does not specify the matrix elements of $T_{ij}$ completely (it may or may not be one of the standard ``random matrix'' forms); we shall however postulate that $T$ and $H_0$ are uncorrelated. It should be carefully noted that the TTLS form of $\chi_{ij:kl}^{(m)}$ is \emph{not} a special case of eqn(\ref{2.4.2}); this may be seen by noting that the form of $Q^{-1}(\omega, T)$ given by the latter is approximately\footnote{Note that because of the high power law in eqn(\ref{2.4.1}), the factor $\theta(E_m+\omega)$ can essentially be neglected in calculating $Q^{-1}(\omega,T)$ (However it is possible that a more refined analysis will lead to a modification of this factor).}   

\begin{align}\label{2.4.3}
 Q^{-1}(\omega,T)=Q^{-1}_0(1-\exp(-\beta\omega))
\end{align}
which is different from eqn(\ref{2.2.12}), though not qualitatively so. What we regard as most important, however, is that the form of $\chi_{ij:kl}^{(m)}(\omega)$ is very different from that which would obtain were all the non-phononic degrees of freedom harmonic oscillators; in the latter case we would simply get the result $Q^{-1}(\omega,T)=\mbox{ind. of } T$, which is qualitatively different from the ``saturating'' forms (\ref{2.4.3}) and (\ref{2.2.12}). Crudely speaking, the ansatz (\ref{2.4.2}) describes a model intermediate between harmonic-oscillator model and the TTLS one, but in some intuitive sense closer to the latter. It is our hope that use of a possibly more realistic form of $\chi^{(m)}$ in the calculations of section 4 will not lead to qualitatively different results.

The expression (\ref{4.1.1}) is finite for all $\omega$ (provided the principal part is correctly taken in the integral). The ratio $R(\omega)$ of the DOS of the interacting superblock to the noninteracting one is given by
\begin{align}\label{4.1.2}
 R(\omega)&=(1+\partial\Delta/\partial\omega)^{-1}\nonumber\\&=(1-CKQ_0^{-2}\ln(U_0/\omega))^{-1}
\end{align}
and in the absence of matrix element renormalization, this should be the expression for $Q^{-1}(\omega)/Q_0^{-1}$. However it is clear that we cannot take this result very seriously, since the expression (\ref{4.1.2}) is evidently negative for small (and even quite large, since the quantity $CKQ_0^{-1}$ is $\gg 1$) values of $\omega$.

What has gone wrong? Let us consider the following heuristic  argument: Suppose that we could somehow neglect all higher-order effects of $\hat{V}$ and thus treat the result (\ref{4.1.1}) as exact. Then a negative value of $R(\omega)$ simply means that the order of levels in the region of $\omega$ has been inverted (i.e. each pair of levels crossed). This still give us a finite level density, but it is now given relative to its original value by the \emph{modulus} of the expression (\ref{4.1.2})! Of course, when we introduce the higher-order effects of $\hat{V}$, we will find that, barring pathology, we do not get level crossing, but rather the familiar level-repulsion effect;  however, if we pretend for the moment that the overall effect of these higher-order terms are ``small'' relative to the second-order ones, the resulting DOS should not be much affected by the lack of crossing. Consequently, we claim that at least over the large regime of $\omega$ for which $CK^2Q_0^{-2}\ln(U_0/\omega)\gg1$, we have
\begin{align}\label{4.1.3}
Q^{-1}(\omega)=[CKQ_0^{-2}\ln(U_0/\omega)-1]^{-1}
\end{align}
which implies that for small $\omega$
\begin{align}\label{4.1.4}
 Q^{-1}(\omega)=\mbox{const.}(\ln U_0/\omega)^{-1}
\end{align}
It is clear that we can extend this result to take into account all pairs $i,j$ in (\ref{2.3.4}), in this case by a direct ``single-shot'' calculation. The result is to simply replace the factor $CK$ in (\ref{4.1.3}), as in (\ref{3.2.12}), by $K_0\log_2(R/L)$:
\begin{align}\label{4.1.5}
Q^{-1}(\omega)&=\frac{Q_0^{-1}}{K_0Q_0^{-2}\log_2(R/L)\ln(U_0/\omega)-1}\\
&\approx\mbox{const.}(\ln(U_0/\omega)) \mbox{\,\,\,\,\,\,\,\,if\,\,\,\,} \omega\ll\omega_0
\end{align}
While this form seems at least qualitatively consistent with the experimental data (see section 5), it is not even approximately universal (since the constant is inversely proportional to $Q_0^{-1}$), and even given a cutoff at $\omega\sim U_0$ does not satisfy (\ref{3.2.12}). Moreover, it is not at all clear that  taking into account higher-order terms in $\hat{V}$ will not change the form of $Q^{-1}$ qualitatively.

At this point, we adopt the following explicitly heuristic tactic: We know directly from ultrasound experiments that for $\omega\ll U_0$ the frequency dependence of $Q^{-1}(\omega)$ is at least approximately consistent with that given in (\ref{4.1.5}); we further know from the thermal conductivity data that for $\omega>U_0$, $Q^{-1}$ must have a much larger value, in fact, of order unity. We thus postulate for the overall frequency-dependence of $Q^{-1}(\omega)$ the ansatz\footnote{One may perhaps object that here (and in (\ref{4.1.7})) $U_0$ should be replaced by the renormalized quantity $U$. However, as we shall see, the conclusions are independent of $U_0$.}
\begin{align}\label{4.1.6}
 Q(\omega)^{-1}(\omega)=\frac{1}{\tilde{Q}_0+A\ln(U_0/\omega)}
\end{align}
where $\tilde{Q}_0$ is not necessarily the same as $Q_0$ but is generally of the same order, and may, like $Q_0$, be system-dependent. We then find the value of $A$ from the requirement that (\ref{4.1.6}) be consistent with (\ref{3.2.12}), i.e. that\footnotemark[\value{footnote}]
\begin{align}\label{4.1.7}
 U_0^{-1}\!\!\!\int_0^{U_0}\!\!\!\!\frac{d\omega}{\tilde{Q}_0+A\ln U_0/\omega}&=[(\bar{Q}_0^{-1})^2+K_0\log_2(R/L)]^{-1}\nonumber\\
&\approx K_0^{-1}[\log_2(R/L)]^{-1}
\end{align}

We see by a change of variables that the LHS of (\ref{4.1.7}) is independent of the cutoff $U_0$ and of the form $A^{-1}F(A^{-1}\tilde{Q}_0)$ where
\begin{align}\label{4.1.8}
 F(\xi)\equiv\int_0^1\frac{dx}{1-\xi^{-1}\log x}&=-e^{\xi}\mbox{Ei}(-\xi)\\
&\approx \ln(\xi^2e^{2\gamma})/2
\end{align}
The approximation is valid for $\xi\ll1$, and $e^{2\gamma}=3.17\ldots$ is the Euler-Macheroni constant. Thus, provided $A^{-1}\tilde{Q}_0\ll1$, we have
\begin{align}\label{4.1.9}
 \bar{Q}^{-1}=A^{-1}\ln(e^{\gamma}\tilde{Q}_0/A)
\end{align}
note that this result is only very weakly system-dependent. If we now put $\tilde{Q}_0\approx1$ and require consistency of (\ref{4.1.9}) with (\ref{3.2.12}), we find that $A\approx350$. Now, setting in (\ref{4.1.6}) $\omega/(2\pi)$ to be of the order $1MHz$ (a typical experimental value), we find
\[Q^{-1}(1MHz)=2.7\times10^{-4}\]
which is precisely the ``typical'' experimental value.

Since the ansatz (\ref{4.1.6}) is only a conjecture, it is interesting to consider the somewhat more general form
\[Q^{-1}(\omega)=(Q_0^{1/s}+A\ln(U_0/\omega))^{-s}\]
 and inquire about the sensitivity of the value of $Q^{-1}(1\mbox{MHz})$ inferred from requiring consistency with (\ref{3.2.12}). For a general $s$ this requires numerical calculation; some representative results are shown in Table-1. 
\begin{table}[h]
\begin{center}
\begin{tabular}{c|c|c|c|c}
  & $s=0.5$ & $s=0.7$ & $s=0.9$\\ \hline
 $Q_0=0.1$ & 0.0024 & 0.0011 & 0.0006\\ \hline
 $Q_0=1$ & 0.0024 & 0.0010 & 0.0004\\ \hline
 $Q_0=10$ & 0.0024 & 0.0010 & 0.0003\\ \hline
\end{tabular}
\caption{The dependence of the value of $Q(\omega\sim1MHz)$ to its functional form}
\end{center}
\end{table}
We see that while the value of $Q^{-1}$ is appreciably sensitive to the (presumably system independent) parameter $s$, it is only very weakly sensitive to the system-dependent quantity $Q_0$; thus, perhaps unsurprisingly in view of the remarks in section 3, the ``universality'' of the ultrasonic absorption is \emph{more general} than the specific ansatz (\ref{4.1.6}).

\section{Discussion}
In section 3 and 4 we have attempted to draw some conclusions concerning the absorption of ultrasound at $T=0$, in the MHz-GHz frequency range. However, no existing ultrasound experiments has ever probed the regime $\omega\gg T$. Thus, in this section we shall compare our predictions with data on related quantities such as the temperature-dependence of the ultrasound absorption and velocity, and the thermal conductivity.  For the purposes of this discussion we will assume (cf section 3) that the many body states with $\omega<U$ are typical of the states as a whole, i.e. that the (renormalized) quantity $\chi_{ijkl}^{(m)}(\omega)$ is ``on average'' independent of $m$, except for the factor $\theta(E_{m}+\omega)$; this greatly simplifies the predictions for the temperature-dependencies of various quantities. 

\subsection{Temperature-Dependence of the Ultrasound Absorption and Velocity}
In the standard TTLS model the temperature-dependence of the absorption, expressed in terms of $Q^{-1}(\omega)$ as above, is given by 
\begin{align}\label{5.1.1}
 Q^{-1}(\omega,T)=Q^{-1}_{hf}\tanh(\omega/2T)
\end{align}
(where $Q^{-1}_{hf}$ is predicted to be independent of $\omega$). In the present model the predicted dependence is straightforwardly calculated by taking into account the possibility of transitions with stimulated emission as well as absorption of a phonon, and using the assumption above concerning $\chi^{(m)}(\omega)$, we find
\begin{align}\label{5.1.2}
Q^{-1}(\omega,T)=Q_{hf}^{-1}(\omega)(1-\mbox{exp}(-\omega/T)) 
\end{align}
where now $Q_{hf}^{-1}=\mbox{const.}/\ln(U/\omega)$. In Figure 1 we show a comparison of the two formulae (\ref{5.1.1}) and (\ref{5.1.2}) with experimental data \cite{golding76} which probes the regime $\omega>T$. Since most experiments, including that of ref. \cite{golding76}, vary $T$ rather than $\omega$, we note that when inferring the quantity $Q_{hf}^{-1}$ (i.e. the value of $Q^{-1}(\omega)$ calculated in section 4) from the raw data, which is actually taken in the regime $\omega\ll T$ we should divide the value of $Q_{hf}$ inferred within the TTLS model by a factor of $2$.

We next turn to the temperature dependence of the ultrasound velocity, $\Delta c(T)$, expressed as a fraction of the velocity $c_0$ at some reference temperature $T_0$\footnote{Because of the very small value of $\Delta c/c$, the precise choice of $T_0$ makes very little difference in the following.}. From the standard Kramers-Kronig relation for the stress-stress response function, we have\footnote{Some care is necessary with the sign; cf. the definition of $\chi(\omega)$ in eqn.(\ref{2.2.10})} 
\begin{align}\label{5.1.3}
 \frac{\Delta c}{c_0}=\frac{-\Delta\chi_0}{\rho c_0^2}=-\Delta\int_0^\infty\frac{d\omega}{\omega}Q^{-1}(\omega,T).
\end{align}
Were we to insert in (\ref{5.1.3}) the TTLS form $Q^{-1}(\omega,T)$, eqn(\ref{5.1.1}), we would get the standard result for $T\gg\omega$
\begin{align}\label{5.1.4}
 \left.\frac{\Delta c}{c_0}\right|_{\mbox{TTLS}}=Q^{-1}_{hf}\ln(T/T_0)
\end{align}
Note that the TTLS requires a modification to its density of states $n(E)=n_0(1+a\omega^2)$ with free parameter $a$ to fit the data above $\sim0.5K$ (cf. \cite{golding76b} and Figure 1).
In the present model, were we to treat $Q^{-1}_{hf}$ as a constant, we would get,
\begin{align}
\Delta c/c= \Delta (Q_{hf}^{-1}/2)[-e^{\omega/T}\mbox{Ei}(-\omega/T)-e^{-\omega/T}\mbox{Ei}(\omega/T)]
\end{align}
which is precisely the same result as (\ref{5.1.4}) in the $\omega\ll T$ limit. In the presence of $[\ln(U_0/\omega)]^{-1}$, the integration must be done numerically, and is shown in Figure 1 and compared with the TTLS result (with unmodified density of states) as well as experimental data\cite{golding76,golding76b}.

\begin{figure}[!ht]
\includegraphics[width=2.5in]{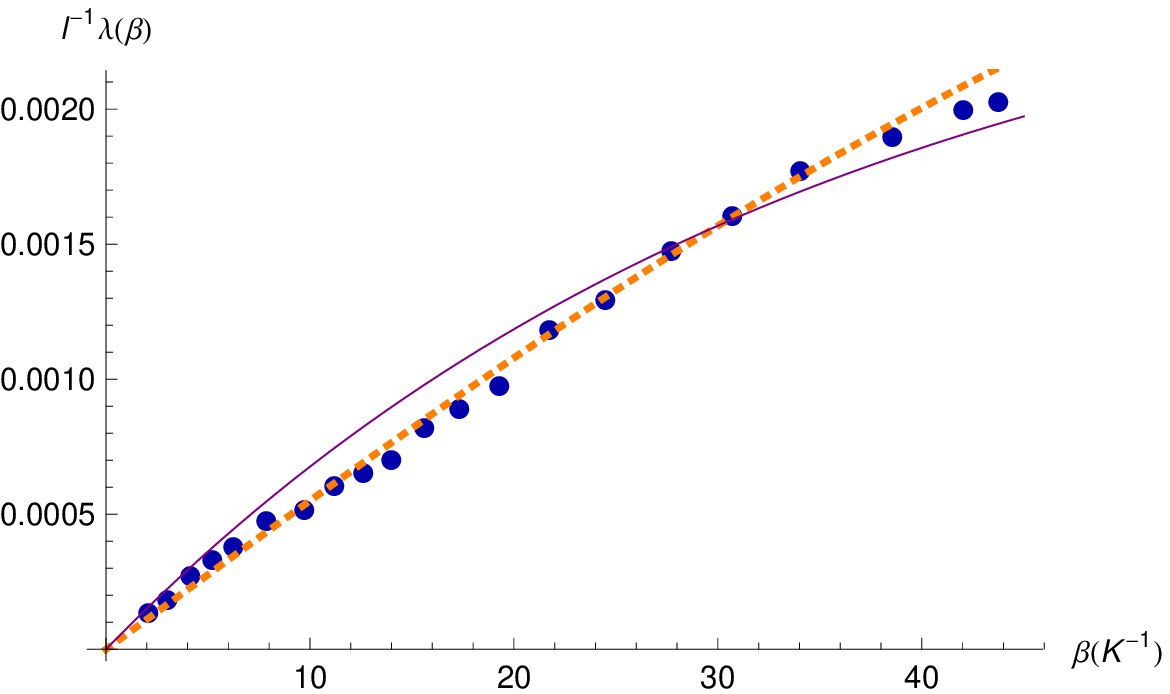}
\includegraphics[width=2.5in]{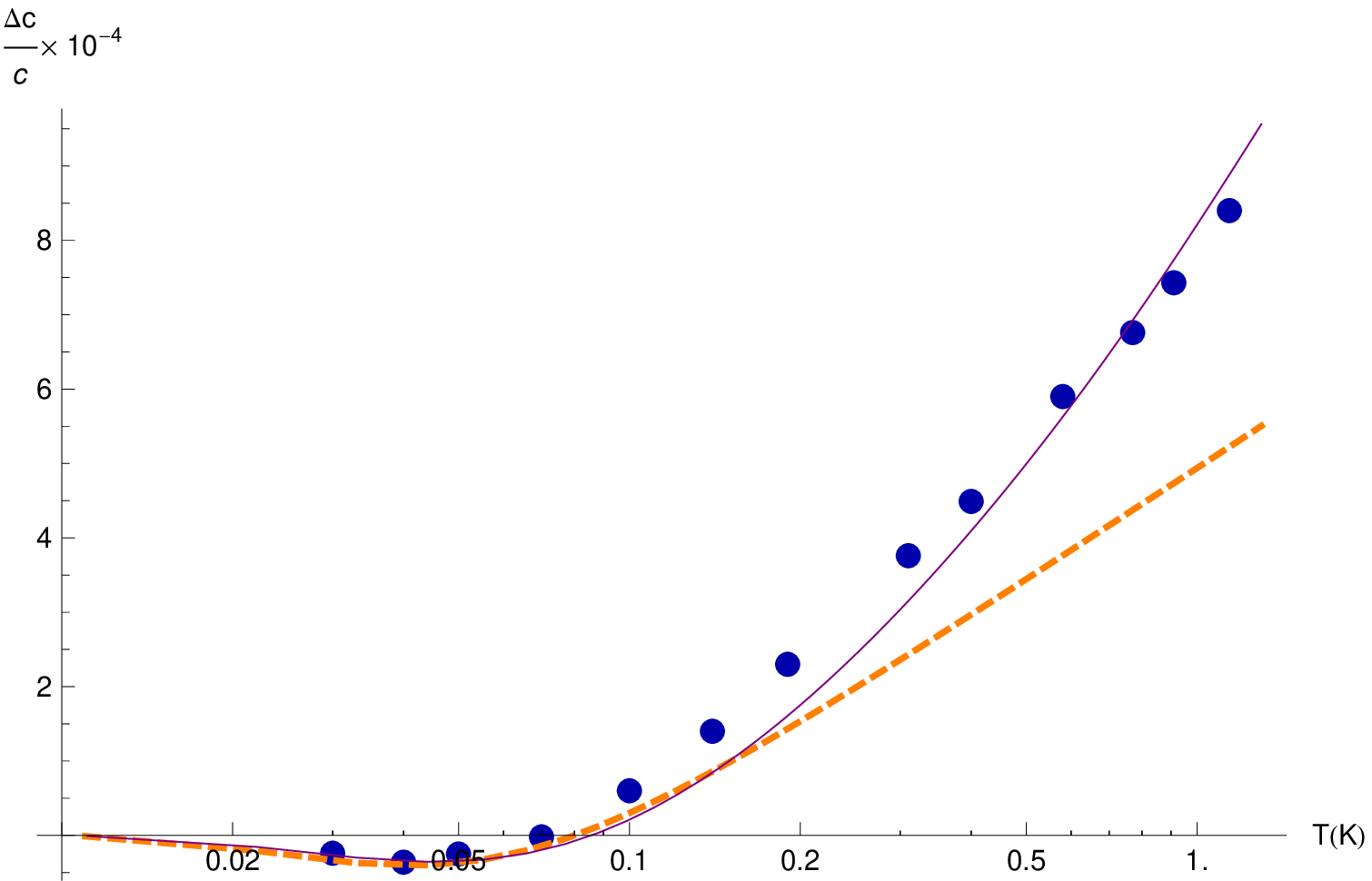}
\caption{
{\bf Normalized inverse mean free path (top) and velocity shift (bottom).} The present theory (solid) is compared against TTLS without the $\omega^2$ term (dashed) and experiment \cite{golding76,golding76b} (dots). While the functional forms predicted by both models are qualitatively similar at low temperatures, the TTLS model must use an additional fitting function $n(\omega)=n_0(1+a\omega^2)$ for the density of states to resolve the discrepancy in fitting $\Delta c/c$ data (bottom).}
\label{Figure1}
\end{figure}

\subsection{Thermal Conductivity}
It is well known \cite{zaitlin75} that the thermal conductivity in amorphous materials below 1K is due entirely to phonons (any other degrees of freedom being effectively localized). If so, then it should be well approximated by the simple kinetic-theory formula
\begin{align}
K=\frac{1}{3}C_v^{ph}\bar{c}\bar{l}_{ph} 
\end{align}
where $C_v^{ph}$ is the phonon contribution to specific heat, $\bar{c}$ is a phonon velocity appropriately averaged over polarization $(l,t)$ and $\bar{l}_{ph}$ is a polarization and frequency-averaged phonon mean free path, which is proportional to $Q$. Since the frequency weighting heavily weights frequencies $\sim4k_BT$, we may set $Q(\omega,T)$ approximately equal to its high frequency value and set $\bar{l}_{ph}=\mbox{const.}Q_{hf}(4T)/4T$. Then for the TTLS model, $\bar{l}^{-1}_{ph}\propto T$, so the thermal conductivity $K$ is predicted to be proportional to $T^2$. In the present model, $\bar{l}_{ph}=\ln(U_0/T)/T$, so we predict
\begin{align}\label{5.2.2}
 K=\mbox{const.}T^2\ln(U_0/T)
\end{align}
-a temperature dependence which is consistent with the usually quoted dependence of the experimentally measured thermal conductivity, namely $K\propto T^{2-\beta}$, $\beta\sim0.05-0.2$, and certainly fits it better than the TTLS prediction $\beta=0$ (see Fig.2).
\begin{figure}[!ht]
\includegraphics[width=2.3in]{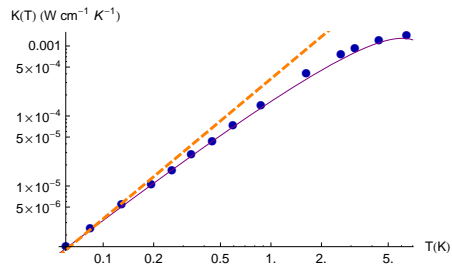}
\caption{
{\bf Temperature Dependence of Thermal Conductivity}. The present theory (solid) $K\sim T^2\ln U/T$ is compared against TTLS prediction $K\sim T^2$ (dashed) and experiment \cite{stephens73} (dots) below the ``plateau''.}
\label{Figure1}
\end{figure}
\section{Conclusion}
As already emphasized, our goal in this paper has been the relatively modest one of investigating how far it is possible to understand the linear ultrasonic properties of glasses in the ``resonant'' regime, and in particular the small and near-universal value of the zero-temperature dimensionless absorption $Q^{-1}(\omega)$, in terms of a model much more generic than the established TTLS model.

We believe that we have been at least partially successful in doing so, in that we have shown that given assumptions (A) and (B) of section 3, a suitable frequency average of $Q(\omega)$ is quasi-universal, in the sense that the only system-specific parameters on which it depends are the ratio $c_l/c_t$ and $\chi_l/\chi_t$ which fluctuate only by a factor of about 1.2 between different materials (see Appendix, and ref\cite{mb}, Fig.1 and 3), and the high-energy cut-off $U_0\sim hc/L$, which enters only logarithmically. (b) that given the ansatz $(\ln\omega)^{-1}$ for the frequency dependence of $Q^{-1}(\omega)$, the absolute value of the absorption is close to the experimental one. (c) that other properties related to $Q^{-1}(\omega)$ (the temperature dependence of $Q^{-1}$ and of the velocity shift, and the thermal conductivity) are consistent with our generic model.

In this paper we have not attempted to discuss two other characteristic properties of glasses which are generally regarded as strong evidence for the TTLS model, namely the nonlinear behavior (saturation, echoes, hole-burning...) and the linear ultrasonic behavior in the so-called ``relaxation regime''. We believe that it is not impossible that a complete renormalization calculation would produce, at the ``output'' stage, a form of $\chi^{(m)}(\omega)$ (suitably averaged over $m$) identical to that following from the TTLS model, in which case it would of course account for all the above phenomena equally well. 

Alternatively, if we assume that the output form of $\chi^{(m)}(\omega)$ is identical to the postulated input form (\ref{4.1.6}) apart from the constant and a factor $(\ln\omega)^{-1}$, it is intuitively obvious that at least some of the nonlinear behavior (e.g. acoustic saturation) will be qualitatively similar to that in the TTLS model although the details will be different. The linear ultrasonic absorption in the low-frequency ``relaxation'' regime is however particularly puzzling, since the successful prediction of the experimentally observed frequency- and temperature-(in)dependence (and even the magnitude) by the TTLS model seems, at least at first sight, to be a consequence of parameter-dependencies very specific to that model. At the time of writing it is not clear to us whether it will be possible to reproduce these dependencies in our more generic model. Of course, an alternative point of view would be that the TTLS spectrum is in fact always realized at the ``input'' stage (i.e. eqn(\ref{2.4.2}) needs to be modified) and that its qualitative features are preserved through the renormalization process; the main role of this process, would then be to give rise to the universal small value of $Q^{-1}$, and the final picture which would then emerge would be somewhat similar to that of Burin and Kagan \cite{burin96}.

Finally,we note that the only property of amorphous materials which we have invoked in this paper is the conjectured existence in them of an appreciable number of non-harmonic degrees of freedom. Thus, our arguments should apply equally well to ``disordered crystals'' (cf. \cite{stamp09}), with the slight modification that the angular dependence of $\Lambda_{ijkl}$ may be different in detail. However, we also note that nothing in our argument implies that \emph{all} amorphous materials (or a fortiori disordered crystals) must have the "canonical" value of $Q^{-1}$; if the input (scale-$r_0$) value of $Q^{-1}$ is appreciably \emph{smaller} than this value, then so will be the output one! However, observation of an amorphous solid with the output $Q^{-1}$ appreciably \emph{greater} than the canonical value would tell against our hypothesis.

This work was supported by the National Science Foundation under grant numbers NSF-DMR-03-50842 and NSF-DMR09-06921. We acknowledge helpful discussions over many years with numerous colleagues, in particular Doug Osheroff, Pragya Shukla, Karin Dahmen, Alexander Burin, Zvi Ovadyahu and Philip Stamp. The first author would like to thank University of British Columbia, University of Waterloo, and Harvard University for generously providing accommodation during long visits.

 \appendix
\section{Transverse to Longitudinal Elastic Ratios}

The present model depends on the numerical values of two near-universal ``inputs'', $c_t/c_l$ and $\mbox{Im}\chi_t/\mbox{Im}\chi_l$. While these values are experimentally justified \cite{mb}, we feel it may be instructive to add some theoretical justification at least to the trivial first ratio. 

Consider a many-body Hamiltonian that consists of an arbitrary inter-block interaction $\phi(r)$ and kinetic energy, which for small strains reduces to (\ref{2.2.1})\footnote{According to the virial theorem the expectation value of the potential energy is equal to that of the kinetic energy for a harmonic Hamiltonian for both longitudinal and transverse terms; thus considering the zero point kinetic energies does not change the ratio $\chi_{0t}/\chi_{0t}$ and will be omitted from our notation.}
\begin{align}
 \hat{H}_{el}(e_{ij})=\mbox{const}+L^{-3}\int \hat{\phi}(r)dr^3
\end{align}
Further, suppose that the relative displacement $u_{ij}$ of two blocks $i$ and $j$ are proportional to the distance $r_{ij}$ between them. For example for purely longitudinal and transverse strains, we would have $u_{x}=e_{xy}r_{y}$ and $u_{x}=e_{xx}r_{x}$ respectively. Then, by definition,
\begin{align}\label{mb-main}
\chi_{0t,l}=\left.\frac{\partial^2}{\partial e_{ij}^2}L^{-3}\int\langle\phi(r)\rangle d^3r\right|_{e_{ij}=0}
\end{align}
where $i\neq j$ and $i=j$ give $\chi_{0t}$ and $\chi_{0l}$ respectively. Letting $\partial\phi/\partial r=0$ due to stability, and taking the derivative $\partial/\partial e_{ij}$ using the distance proportionality assumption, we obtain $\chi_{0t}\propto \int r_x^2r_y^2 d\Omega$ and $\chi_{0l}\propto\int r_x^4 d\Omega$, where $\Omega$ is the solid angle, and the constants of proportionality are identical. Since the speed of sound is related to the real part of the zero frequency response function according to (\ref{2.1.6}), we get
\begin{align}
 \frac{c_t}{c_l}=\sqrt{\frac{\chi_{0t}}{\chi_{0l}}}=\frac{1}{\sqrt{3}}
\end{align}
which is $6\%$ larger than the experimental (average) value.

\section{Evaluation of the Traces}

In this appendix we outline the algebra involved in obtaining eqn(\ref{3.2.5}) and eqn(\ref{3.2.8}) (the derivation of eqn(\ref{3.2.6}) is similar to the latter). 

Let us expand the non-interacting and interacting two-block density of states $p(E)$ in powers of $E$, such that the the coefficient of the $n^{th}$ power is $c_{0n}$ and $c_{n}$ respectively (cf. eqn(\ref{3.2.3})). For each set of coefficients the normalization conditions require that the two body system has $N_s^2$ levels whether they interact or not
\begin{eqnarray}\label{N}
\sum_{n=0}^{p}c_{0n}\frac{U_0^{n_0+1}}{n_0+1}=\sum_{n=0}^{p}c_n\frac{U^{n+1}}{n+1}=N_s^2\nonumber\\
\end{eqnarray}
The trace in question is
\[\mbox{Tr}H^2=\int_0^U\omega^2f(\omega)d\omega=\sum_{n=0}c_n\frac{U^{n+3}}{n+3},\]
if the terms for which $n\gg1$ dominate the density of states (a premise required for the extensivity of many-body energy levels), we can use eqn(\ref{N}) to write
\begin{eqnarray}\label{dosintegral}
\mbox{Tr}H^2=U^2\sum_{n=0}c_n\frac{U^{n+1}}{n+3}\approx U^2N_s^2.
\end{eqnarray}
The evaluation of $\mbox{Tr}H_0^2$ is similar. Now let us turn to the trace of the square of
\begin{align}
\hat{V}_{ab}=\frac{2}{3v_{ab}\rho c_t^2}\tilde{\Lambda}_{ijkl}\hat{T}_{ij}^{(a)}\hat{T}_{kl}^{(b)}
\end{align}
where $v_{ab}=4\pi r_{ab}^3/3$ is the volume of a sphere defined by the inter-block separation $r_{ab}$.
\begin{align}
M=Tr V^2=\sum_{\substack{n_1n_2\\m_1m_2}}\sum_{\substack{ijkl\\i'j'k'l'}}\Lambda_{ijkl}\Lambda_{i'j'k'l'}\nonumber\\
\times\langle n_1|\hat{T}_{ij}^{a}|m_1\rangle\langle n_1|\hat{T}_{i'j'}^{a}|m_1\rangle\langle n_2|\hat{T}_{kl}^{b}|m_2\rangle\langle n_2|\hat{T}_{k'l'}^{b}|m_2\rangle 
\end{align}

where $|n_s\rangle$ is the $n^{th}$ eigenstate of block $s$. The sum over states $n_s$ and $m_s$ is over the manifold described in the text, such that $0<E_{m_s},E_{n_s}<U$. Thus, for all $n_s$ we may write the identity as
\[\int_{-E_{m_s}}^{U-E_{m_s}}\delta(E_{n_s}-E_{m_s}-\omega')d\omega,\]
insert it in the sum twice with $s=1,2$ and evaluate the sums over $n_1,n_2$
\begin{align}
M=\frac{4v_av_b}{9(\pi v_{ab}\rho c_t^2)^2}\sum_{m_1m_2}\int_{-E_{m_1}}^{U-E_{m_1}}\int_{-E_{m_2}}^{U-E_{m_2}} d\omega'd\omega''\nonumber\\
\times\sum_{\substack{ijkl\\i'j'k'l'}}\tilde{\Lambda}_{ijkl}\tilde{\Lambda}_{i'j'k'l'}\chi^{(a)}_{ij:i'j'}(\omega')\chi^{(b)}_{kl:k'l'}(\omega'')\\
 =\frac{4v_av_b}{9(\pi v_{ab}\rho c_t^2)^2}U^2N_s^2\sum_{\substack{ijkl\\i'j'k'l'}}\Lambda_{ijkl}\Lambda_{i'j'k'l'}\bar{\chi}_{ij:i'j'}^{a}\bar{\chi}_{kl:k'l'}^{b}
\end{align}

Now let us decompose each factor of $\bar{\chi}_{ij:kl}$ into the only two independent components, as in eqn(\ref{2.1.5}). Since $\tilde{\Lambda}_{ijkl}$ only depends on $q$, and any $\chi_{ij:kl}$ can be written in terms of $\bar{\chi}_t$ and $p$, it is obvious that the right hand side will be proportional to $\left(\bar{Q}_t^{-1}\right)^2$. 

\begin{align}
 M= \frac{v_av_b}{v_{ab}}K\left(\bar{Q}_t^{-1}\right)^2
\end{align}
where 
\begin{align}
K\equiv\frac{4}{9\bar{\chi}_t^2}\sum_{\substack{ijkl\\i'j'k'l'}}\Lambda_{ijkl}\Lambda_{i'j'k'l'}\bar{\chi}_{ij:i'j'}\bar{\chi}_{kl:k'l'}
\end{align}

While the evaluation of the sum over all tensor components may seem complicated at first sight, the symmetries of $\chi_{iji'j'}\chi_{klk'l'}$ simplify the problem considerably. Any term including a $\chi_{ijkl}$ with odd number of equal indices (such as $\chi_{1323}$ or $\chi_{2223}$) vanish. Furthermore, any term with indices $\{ijkli'j'k'l'\}$ is equal to that with $\{i'j'klijk'l'\}$, $\{ijk'l'ijkl\}$ and $\{i'j'k'l'ijkl\}$. The sum is evaluated to yield $K=(8/3) \left(-3+16 (-1+q) q+4 p (1+2 q)^2\right)\approx122$.

\bibliographystyle{unsrt}
\bibliography{mybib}

\begin{thebibliography}{10}

\bibitem{pohl71}
R.C Zeller and R.O Pohl.
\newblock Thermal conductivity and specific heat of noncrystalline solids.
\newblock {\em Physical Review B}, 4(6):2029, 1971.

\bibitem{anderson86}
A.C. Anderson and J.J. Freeman.
\newblock Thermal conductivity of amorphous solids.
\newblock {\em Physical Review B}, 34(8):5684--5690, 1986.

\bibitem{mb}
M.~Meissner and J.F. Berret.
\newblock How universal are the low-temperature acoustic properties of glasses.
\newblock {\em Zeitschrift fur Physik B}, 70(1):65--72, 1988.

\bibitem{pohl2002}
R.O. Pohl, X.~Liu, and E.~Thompson.
\newblock Low-temperature thermal conductivity and acoustic attenuation in
  amorphous solids.
\newblock {\em Reviews of Modern Physics}, 74(4):991--1013, 2002.

\bibitem{deyoreo86}
J.J. {De Yoreo}, W.~Knaak, M.~Meissner, and R.O. Pohl.
\newblock Low temperature properties of crystalline (kbr)\_{1-x}(kcn)\_x: A
  model glass.
\newblock {\em Physical Review B}, 34(12):8828--8842, 1986.

\bibitem{joffrin75}
J.~Joffrin and A.~Levelut.
\newblock Virtual phonon exchange in glasses.
\newblock {\em Journal de Physique}, 37(3):271--273, 1976.

\bibitem{burin96}
A.L. Burin and Y.~Kagan.
\newblock On the nature of universal properties of amorphous solids.
\newblock {\em Physics Letters A}, 215(3-4):191--196, 1996.

\bibitem{leggett88}
A.J Leggett and C.C Yu.
\newblock Low temperature properties of amorphous materials: Through a glass
  darkly.
\newblock {\em Comments on Condensed Matter Physics}, 14(4):231--251, 1988.

\bibitem{leggett91}
A.J. Leggett.
\newblock Amorphous materials at low temperatures: Why are they so similar.
\newblock {\em Physica B}, 169(1-4), 1991.

\bibitem{yu89}
C.C. Yu.
\newblock Interacting defect model of glasses - why do phonons go so far.
\newblock {\em Physical Review Letters}, 63(11):1160--1163, 1989.

\bibitem{wolynes01}
V.~Lubchenko and P.G. Wolyness.
\newblock Intrinsic quantum excitations of low temperature glasses.
\newblock {\em Physical Review Letters}, 87(19), 2001.

\bibitem{turlakov04}
M.~Turlakov.
\newblock Universal sound absorption in low-temperature amorphous solids.
\newblock {\em Physical Review Letters}, 93(3), 2004.

\bibitem{parshin94}
D.A. Parshin.
\newblock Interactions of soft atomic potentials and universality of
  low-temperature properties of glasses.
\newblock {\em Physical Review B}, 49:9400--9418, 1994.

\bibitem{stamp09}
M.~Schechter and P.C.E Stamp.
\newblock Low-temperature universality in disordered solids.
\newblock (arXiv: 0910.1283v1), 2009.

\bibitem{pines}
P.~Nozieres and D.~Pines.
\newblock {\em Theory of Quantum Liquids}.
\newblock Benjamine, New York, 1966.

\bibitem{esquinazi-rev}
A.L. Burin, D.~Natelson, D.D. Osheroff, and Y.~Kagan.
\newblock {\em Tunneling Systems in Amorphous and Crystalline Solids}.
\newblock Springer-Verlag, 1998.

\bibitem{fisch80}
R.~Fisch.
\newblock Scaling theory of glass.
\newblock {\em Physical Review B}, 22(7):3459--3464, 1980.

\bibitem{dcv-future}
D.C. Vural.
\newblock Unpublished.

\bibitem{golding76}
B.~Golding, J.E. Graebner, and R.J. Schutz.
\newblock Intrinsic decay lengths of quasimonochromatic phonons in a glass
  below 1k.
\newblock {\em Physical Review B}, 14(4):1660--1662, 1976.

\bibitem{golding76b}
B.~Golding, J.E. Graebner, and A.B. Kane.
\newblock Phase velocity of high-frequency phonons in glasses below 1k.
\newblock {\em Physical Review Letters}, 37(18):1248--1250, 1976.

\bibitem{zaitlin75}
M.~P. Zaitlin and A.C. Anderson.
\newblock Phonon thermal transport in noncrystalline materials.
\newblock {\em Physical Review B}, 12(10):4475--4487, 1075.

\bibitem{stephens73}
R.B. Stephens.
\newblock Low-temperature specific heat and thermal conductivity of
  noncrystalline dielectric solids.
\newblock {\em Physical Review B}, 8(6):2896--2905, 1973.

\end{thebibliography}







\end{document}